\documentclass{elsart}

\usepackage{amssymb}
\usepackage{graphicx}
\usepackage{amsmath}

\begin{document}

\begin{frontmatter}



\title{Design of Electromagnetic Cloaks and Concentrators Using Form-Invariant
Coordinate Transformations of Maxwell's Equations}


\author{Marco Rahm,}
\author{David Schurig,}
\author{Daniel A. Roberts,}
\author{Steven A. Cummer,}
\author{David R. Smith}
\address{Department of Electrical and Computer Engineering,
Duke University, Box 90291, Durham, NC 27708, USA}
\author{Sir John B. Pendry}
\address{Department
of Physics, The Blackett Laboratory, Imperial College, London SW7
2AZ, UK}


\begin{abstract}
The technique of applying form-invariant, spatial coordinate
transformations of Maxwell's equations can facilitate the design of
structures with unique electromagnetic or optical functionality.
Here, we illustrate the transformation-optical approach in the
designs of a square electromagnetic cloak and an omni-directional
electromagnetic field concentrator.  The transformation equations
are described and the functionality of the devices is numerically
confirmed by two-dimensional finite element simulations. The two
devices presented demonstrate that the transformation optic approach
leads to the specification of complex, anisotropic and inhomogeneous
materials with well directed and distinct electromagnetic behavior.
\end{abstract}

\end{frontmatter}

\section{Introduction}
\label{sec:Introduction}
In a theoretical study, Pendry et al. reported a general method for
the design of electromagnetic materials based on form-invariant
transformations of Maxwell's equations \cite{Pendry:2006}. In that
paper, the methodology of transformation optics was applied to find
the specification for an electromagnetic cloak-a complex material
capable of rendering objects within its interior invisible to
detection. Although just one example of the many intriguing
structures possible using the transformation optical approach, the
proposed cloak design generated enormous interest in its own right.
An approximation to the invisibility cloak based on metamaterials
was subsequently realized by Schurig et al., who demonstrated the
cloaking mechanism in microwave experiments
\cite{SchurigScience:2006}. The transformation optical approach to
invisibility is quite general, differing in scope from prior related
work. Indeed, methods of reducing the electromagnetic scattering of
objects at radar frequencies have long been a subject of intense
research \cite{Kay:1965,Rusch:1976,Kildal:1996}. On the nanoscale,
techniques have also been suggested to reduce the scattering of one
or more multipole components of size-limited objects using tailored
negative index or negative permittivity coatings \cite{Engheta:2005,
Kerker:1975}. More recently, a mathematically rigorous proof of an
invisibility structure based on active devices was reported
\cite{Greenleaf:2007}.

Transformation optics provides for a conceptually simple approach to
the design of complex electromagnetic structures: one imagines
warping space to achieve the desired electromagnetic functionality.
The trajectories of electromagnetic waves passing through a region
of warped space must conform to the local metric, and this provides
an alternative (though conceptual) means to control and manipulate
electromagnetic fields. Once the desired design is determined, the
coordinate transformation and its Jacobi matrix determine the
transformation of  Maxwell's equations and the constitutive
relations. The result provides the specification for an
electromagnetic structure that is complex-being inhomogeneous and
anisotropic-but realizable for example through artificially
structured metamaterials. Indeed, because the fields in a volume
bounding a transformation optical structure are identical to those
that would exist where the structure is replaced by free space,
anisotropy is necessary to circumvent uniqueness constraints
\cite{Greenleaf:2007}.

If the coordinate transformation can be realized exactly in the
constitutive parameters, all aspects of wave propagation will be
transformed by the structure, including the near-fields. Adding
constraints to the materials reduces the ultimate performance of the
structure, but nevertheless can still allow for interesting and
novel structures. Leonhardt, for example, has shown that if the
materials are restricted to be isotropic, an approximate cloak can
be constructed that is valid in the geometrical optic limit
\cite{Leonhardt:2006}. Likewise, constraints were employed for the
metamaterial cloak utilized by Schurig et al.\ to ease the
metamaterial design and fabrication, resulting in a structure that
produced significant reflection yet still demonstrated the cloaking
mechanism for transmitted waves \cite{SchurigScience:2006}.

Since the concept of transformation optics was introduced, there
have been a growing number of subsequent reports applying the method
to a variety of electromagnetic, acoustic and elasto-mechanical
structures
\cite{Milton1:2006,Milton2:2006,Cummer:2007,Shalaev:2007,Ruan:2007,Chen:2007,Miller:2006,Wood:2007}.
Full wave simulations have helped to confirm the expected behavior
and have provided a platform to explore systematically the effects
of absorption, imperfections and other constraints that are inherent
to fabricated realizations of the transformation optical structures
\cite{Cummer:2006,Zolla:2007}.

In this paper we present two examples that demonstrate the general
applicability of form-invariant coordinate transformations for the
design of complex, inhomogeneous and anisotropic electromagnetic
materials with well-defined functionality. For the first example, we
derive the electromagnetic constitutive parameters corresponding to
a two-dimensional electromagnetic cloak having square cross-section.
The square shape has been chosen to illustrate the nature of the
transformation-and the resulting design-for a structure that lacks
rotational symmetry in the plane. In contrast to the cylindrical
cloaks with circular cross-section previously presented, the square
cloak design results in a non-orthogonal transformation producing a
more complicated specification for the spatially dependent
permittivity and permeability tensors. The method to design this
structure, however, can be applied to the design of structures with
arbitrary shape.

For the second example, we derive the material properties of an
electromagnetic field concentrator by the same approach. The purpose
of the cylindrical concentrator is to focus incident electromagnetic
waves with wave vectors perpendicular to the cylinder axis,
enhancing the electromagnetic energy density of incident waves in a
given area. This example illustrates the strength of the
transformation-optical approach for designing devices other than
cloaks.
\section{Transformation Equations}
\label{sec:Trafos}
In this section, the formulas describing the spatial coordinate
transformations and the calculation of the resulting material
parameters, i.\ e.\ the electric permittivity tensor and the
magnetic permeability tensor, are derived. The methodology used to
compute the electromagnetic material properties is similar to the
one reported in \cite{Schurig:2006}.

For convenience, we denote Maxwell's equations in the Minkowski form
\cite{Post:1997}
\begin{eqnarray}
\partial_{[\kappa}F_{\lambda \nu ]} & = & 0 \label{formula:Mink1} \\
\partial_{\nu}G^{\nu \lambda} & = & j^{\lambda}, \qquad
\kappa,\lambda,\nu = 0,1,2,3 \label{formula:Mink2}
\end{eqnarray}
where the square brackets express an alternation among the indices
\cite{Schouten:1951} and the skew-symmetric covariant and
contravariant tensors $F_{\lambda \nu}$ and $G^{\nu \lambda}$ and
the contravariant
 vector $j^{\lambda}$ possess the identifications
\begin{eqnarray}
F_{\lambda \nu } & = & \left( \begin{array}{cccc} 0 & -E_{1} &
-E_{2} & -E_{3} \\ E_{1} & 0 & B_{3} & -B_{2} \\ E_{2} & -B_{3}
 & 0 & B_{1} \\ E_{3} & B_{2} & -B_{1} & 0 \end{array} \right) \\
G^{\nu \lambda} & = & \left( \begin{array}{cccc} 0 & D_{1} & D_{2} &
D_{3} \\ -D_{1} & 0 & H_{3} & -H_{2} \\ -D_{2} & -H_{3}
 & 0 & H_{1} \\ -D_{3} & H_{2} & -H_{1} & 0 \end{array} \right) \\
 j^{\lambda} & = & \left( \begin{array}{c} \rho \\ j_{1} \\j_{2}
 \\j_{3} \end{array} \right)
\end{eqnarray}
where $E$ is the electric field, $B$ is the magnetic induction, $D$
is the electric displacement, $H$ is the magnetic field, $\rho$ is
the volume charge density, $j$ is the current density and the
indices denote their spatial components. In this notation, the
coordinate vector in four-space is $x^{\alpha} = (x_{0} = t, x_{1},
x_{2}, x_{3})^{T}$ with the vacuum speed of light $c$ set to unity.

For a linear medium, the constitutive relation can be written as
\begin{equation}
G^{\lambda \nu} = \frac{1}{2} \chi^{\lambda \nu \sigma \kappa}
F_{\sigma \kappa} \label{formula:constitutive}
\end{equation}
where the tensor $\chi^{\lambda \nu \sigma \kappa}$ contains the
complete information about the electromagnetic material properties.

The Minkowski equations (\ref{formula:Mink1}) and
(\ref{formula:Mink2}) and the constitutive relation
(\ref{formula:constitutive}) are form-invariant for arbitrary
continuous space-time transformations of the form
\begin{equation}
x^{\alpha'}(x^{\alpha}) =  A^{\alpha'}_{\alpha} x^{\alpha}
\label{formula:xcoordtrafo}
\end{equation}
where $A^{\alpha'}_{\alpha} =
\frac{\partial{x^{\alpha'}}}{\partial{x^{\alpha}}}$ are the elements
of the Jacobian transformation matrix and the primed indices denote
the space-time coordinates of the vector x in the transformed space.

Considering the transformation (\ref{formula:xcoordtrafo}), the
Minkowski equations (\ref{formula:Mink1}) and (\ref{formula:Mink2})
transform as
\begin{eqnarray}
\partial_{[\kappa'}F_{\lambda' \nu' ]}&=& A^{\kappa}_{\kappa'}A^{\lambda}_{\lambda'}A^{\nu}_{\nu'}
\partial_{[\kappa}F_{\lambda \nu ]} = 0 \label{formula:Mink1trafo}
\\
\partial_{\nu'}G^{\nu' \lambda'} &=&
[det(A^{\lambda'}_{\lambda})]^{-1}A^{\lambda'}_{\lambda}
\partial_{\nu} G^{\nu \lambda} \nonumber
\\
&=&[det(A^{\lambda'}_{\lambda})]^{-1}A^{\lambda'}_{\lambda}j^{\lambda}
\\
&=& j^{\lambda'}\label{formula:Mink2trafo}
\end{eqnarray}
and the constitutive relation transforms as
\begin{eqnarray}
G^{\lambda' \nu'} &=&
[det(A^{\lambda'}_{\lambda})]^{-1}A^{\lambda'}_{\lambda}A^{\nu'}_{\nu}G^{\lambda
\nu} \nonumber
\\
&=&
\frac{1}{2}[det(A^{\lambda'}_{\lambda})]^{-1}A^{\lambda'}_{\lambda}A^{\nu'}_{\nu}
A^{\sigma'}_{\sigma}A^{\sigma}_{\sigma'}A^{\kappa'}_{\kappa}A^{\kappa}_{\kappa'}\chi^{\lambda\nu
\sigma \kappa}F_{\sigma \kappa} \nonumber
\\
&=&
\frac{1}{2}[det(A^{\lambda'}_{\lambda})]^{-1}A^{\lambda'}_{\lambda}A^{\nu'}_{\nu}
A^{\sigma'}_{\sigma}A^{\kappa'}_{\kappa}\chi^{\lambda\nu \sigma
\kappa} A^{\sigma}_{\sigma'}A^{\kappa}_{\kappa'}F_{\sigma \kappa}
\nonumber
\\
&=&\frac{1}{2}\chi^{\lambda'\nu' \sigma' \kappa'}F_{\sigma'
\kappa'}\label{formula:constittrafo}
\end{eqnarray}
with
\begin{eqnarray}
\chi^{\lambda'\nu' \sigma' \kappa'} &=&
[det(A^{\lambda'}_{\lambda})]^{-1}A^{\lambda'}_{\lambda}A^{\nu'}_{\nu}
A^{\sigma'}_{\sigma}A^{\kappa'}_{\kappa}\chi^{\lambda\nu \sigma
\kappa} \\
F_{\sigma' \kappa'} &=&
A^{\sigma}_{\sigma'}A^{\kappa}_{\kappa'}F_{\sigma \kappa}
\end{eqnarray}
where $det(X)$ indicates the determinant of a tensor $X$.

It is obvious, that the form-invariance of the Minkowski equations
(\ref{formula:Mink1}) and (\ref{formula:Mink2}) and the constitutive
relation (\ref{formula:constitutive}) also holds for transformations
which only address the space coordinates, as the space manifold is a
sub-manifold of the space-time manifold. In the following we will
restrict ourselves to time-independent, spatial coordinate
transformations. Under this restriction, the constitutive
parameters, i.\ e.\ the tensors of the permittivity and the
permeability of a linear, anisotropic, non-dispersive,
non-bianisotropic and locally interacting medium can be written in a
more accessible form as
\begin{eqnarray}
\epsilon^{i' j'} &=&
[det(A^{i'}_{i})]^{-1}A^{i'}_{i}A^{j'}_{j}\epsilon^{i j}
\label{formula:epsilon} \\
\mu^{i' j'} &=& [det(A^{i'}_{i})]^{-1}A^{i'}_{i} A^{j'}_{j}\mu^{i
j}\label{formula:mu}
\end{eqnarray}
The relations (\ref{formula:xcoordtrafo}), (\ref{formula:epsilon})
and (\ref{formula:mu}) form the underlying equations for the
calculation of the electromagnetic material parameters used in the
design of a square-shaped cloak and a concentrator for
electromagnetic fields.

For all the transformations considered in the next sections, the
mathematical starting point is three-dimensional, euclidian space
expressed in a cartesian coordinate system $x^{i} =
(x_1,x_2,x_3)^{T}$. From the physical point of view, the space is
considered to be medium-free (vacuum) and isotropic. Thus, the
permittivity tensor $\epsilon^{i j}$ and the permeability tensor
$\mu^{i j}$ of the original space can be expressed in the form
\begin{eqnarray}
\epsilon^{i j} &=& \varepsilon_{0} \delta^{i j} \label{formula:epsisotropic} \\
\mu^{i j} &=& \mu_{0}\delta^{i j} \label{formula:muisotropic}
\end{eqnarray}
with ($\delta^{i j}=1$) for ($i=j$) and ($\delta^{i j}=0$)
elsewhere.
\subsection{Square Cloak}
\label{subsec:Trafo_Sqcloak}
\begin{figure}
\begin{center}
\includegraphics[width = 3in]{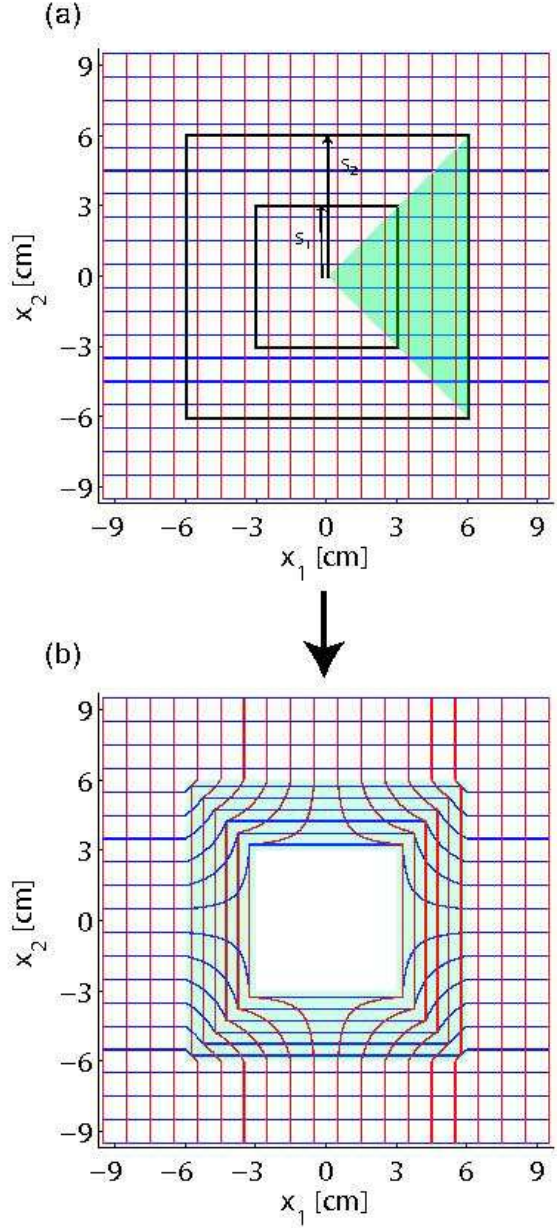}
\caption{\label{fig:SqCloakTrafo} Spatial coordinate transformation
for the design of a square-shaped cloak (a) original space, $s_1$:
half sidelength of the inner square, $s_2$: half sidelength of the
outer square. The transformation equations
(\ref{formula:SqTrafo1})-(\ref{formula:SqTrafo3}) are only valid in
the green shadowed region (b) transformed space}
\end{center}
\end{figure}

The coordinate transformation equations for the electromagnetic
design of a square-shaped cloak with a sidelength $2s_1$ of the
inner square and a sidelength $2s_2$ of the outer square (see fig.\
\ref{fig:SqCloakTrafo}a) are expressed by
\begin{eqnarray}
x_{1}'(x_1,x_2,x_3) &=&  x_{1}\frac{s_2-s_1}{s_2}+s_1 \label{formula:SqTrafo1}\\
x_{2}'(x_1,x_2,x_3) &=&  x_{2} \left( \frac{s_2-s_1}{s_2}+\frac{s_1}{x_1} \right) \label{formula:SqTrafo2} \\
x_{3}'(x_1,x_2,x_3) &=&  x_3 \label{formula:SqTrafo3}
\end{eqnarray}
with the Jacobi matrix and its determinant
\begin{eqnarray}
A^{i'}_{i} &=& \left( \begin{array}{ccc} \frac{s_2-s_1}{s_2} & 0 & 0 \\
\frac{-x_2}{x^{2}_{1}} s_1 & \frac{s_2-s_1}{s_2}+\frac{s_1}{x_1} & 0 \\
0 & 0 & 1 \end{array} \right) \label{formula:JacobiSqCloak} \\
det(A^{i'}_{i}) &=& \frac{s_2-s_1}{s_2}
\left(\frac{s_2-s_1}{s_2}+\frac{s_1}{x_1} \right)
\label{formula:DetJacobiSqCloak}
\end{eqnarray}
for ($0<x_1\leq s_2$), ($-s_2<x_2\leq s_2$), $|x_2|<|x_1|$ and
($|x_3|<\infty$). It should be noted, that, by the foregoing
definitions, the transformation equations
(\ref{formula:SqTrafo1})-(\ref{formula:SqTrafo3}) are only defined
for the green shadowed area in fig.\ \ref{fig:SqCloakTrafo}a and
that the transformation is continuous at the boundary of the
transformed domain. The corresponding transformation formulas for
the upper, left, and lower domain of the square cloak can be readily
obtained by applying rotation operators with rotation angles of $\pi
/2$, $\pi$ and $3\pi /2$ around the $z$-axis to equations
(\ref{formula:SqTrafo1})-(\ref{formula:SqTrafo3}).

As can be seen from fig.\ \ref{fig:SqCloakTrafo}b, the
transformation expands the space within the inner square at the
expense of a compression of space between the inner and outer
square.
\begin{figure*}
\centering
\includegraphics[width=6.5in]{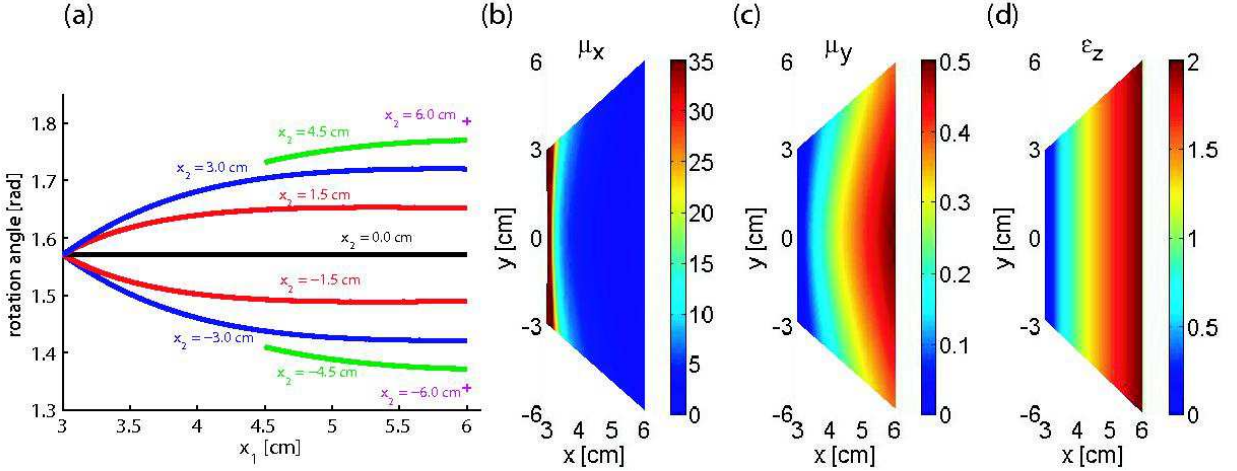}
\caption{\label{fig:SqCloakrotgrid} (a) Rotation angle of the
eigenbasis vectors $(x,y)$ in dependence on their spatial location
$(x_1,x_2)$ within the cloaking material calculated for the green
shadowed region in fig.\ \ref{fig:SqCloakTrafo}a. (b-d) Spatial
dependence of the material parameters of the square cloak medium for
a TE-wave polarized in z-direction, expressed in their local
eigenbasis $(x,y,z)$, (b) permeability $\mu_x$ in x-direction, (c)
permeability $\mu_y$ in y-direction, (d) permittivity in
$\epsilon_z$ in z-direction}
\end{figure*}
Inserting equations (\ref{formula:JacobiSqCloak}-
\ref{formula:DetJacobiSqCloak}),
(\ref{formula:epsisotropic}-\ref{formula:muisotropic}) into
(\ref{formula:epsilon}) and (\ref{formula:mu}), applying the reverse
transformations $x^{i}(x^{i'})$ of
(\ref{formula:SqTrafo1})-(\ref{formula:SqTrafo3}) and dividing
(\ref{formula:epsilon}) by $\varepsilon_{0}$ and (\ref{formula:mu})
by $\mu_{0}$  yields the relative permittivity and the relative
permeability tensors $(\epsilon_{r})^{i' j'}$ and $(\mu_{r})^{i'
j'}$, expressed in the coordinates $x^{i'}$ of the transformed
space, as
\begin{eqnarray}
(\epsilon_{r})^{i' j'} = (\mu_{r})^{i' j'}= \left(
\begin{array}{ccc}
\frac{c}{a} & -\frac{b}{a} & 0 \\
-\frac{b}{a} & \frac{a^{2}+b^{2}}{ac} & 0 \\
0 & 0 & ac
\end{array} \right ) \label{formula:epsmuTrafo}
\end{eqnarray}
%
%
%
with
\begin{eqnarray}
a := \frac{s_2}{s_2-s_1}\textrm{, } b :=\frac{x_{2}'}{(x_{1}')^{2}}
\,a s_1 \textrm{, } c:= a \left(  1-\frac{s_1}{x_{1}'} \right)
\end{eqnarray}
Due to the natural invariance of the Minkowski equations (as
discussed in \cite{Schurig:2006}), the permittivity and permeability
tensors can also be interpreted as the material properties of a
medium described in the coordinate system of the original space by
substituting the primed indices by unprimed indices (''material
interpretation''). Again it should be noted, that the material
properties (\ref{formula:epsmuTrafo}) are only valid in the green
shadowed region of fig.\ \ref{fig:SqCloakTrafo}a and that, due to
the symmetry of the cloak, the material properties of the other
cloak domains can be readily obtained by rotating the tensors in
(\ref{formula:epsmuTrafo}) by $\pi /2$, $\pi$ and $3\pi /2$,
respectively. Furthermore, the relative permittivity and
permeability tensors are non-diagonal, which is a direct consequence
of the non-conformality of the transformation
(\ref{formula:SqTrafo1})-(\ref{formula:SqTrafo3}). However, in terms
of fabricating such a material, it is desirable to have the material
parameters denoted in their eigenbasis, where the permittivity and
permeability tensors are diagonal. Due to the symmetry of the
tensors $\epsilon^{ij}$ and $\mu^{ij}$, an eigenbasis solution
always exists and one obtains
\begin{eqnarray}
(\epsilon_{r})^{i j} = (\mu_{r})^{i j}= \nonumber
\end{eqnarray}
\begin{eqnarray}
\frac{1}{2ac} \left(
\begin{array}{ccc}
 \left( A + \sqrt(B) \right) & 0 & 0 \\
0 & \left(A - \sqrt(B) \right)& 0 \\
0 & 0 & 2a^2c^2
\end{array} \right ) \label{formula:epsmudiag}
\end{eqnarray}
with
\begin{eqnarray}
A &:=& a^2+b^2+c^2 \\
B &:=& a^2\left( a^2+2(b^2-c^2)\right)+b^2(b^2+2c^2)+c^4
\end{eqnarray}

Notice, that the primes were omitted to express the electromagnetic
parameters in the material interpretation. Due to the spatial
dependence of the elements of the permittivity and permeability
tensors, the orientation of the basis vectors $(x,y)$ of the
eigenbasis depends on the spatial location $(x_1,x_2)$ within the
cloak material. This is illustrated in fig.\
\ref{fig:SqCloakrotgrid}a for a square cloak with $s_1=3$~cm and
$s_2=6$~cm. The graph shows the rotation angle of the eigenbasis
vectors $(x,y)$ in dependence on the location $x_1$ at different
positions $x_2$. Again, the physical quantities are calculated for
the green shadowed region in fig.\ \ref{fig:SqCloakTrafo}a. The
rotation angle of the eigenbasis with refer to the coordinate system
$(x_1,x_2)$ varies within a range from 1.3 to 1.8 rad. In order to
fabricate such a medium as a metamaterial, the principle axes of the
unit cells have to be individually aligned along the basis vectors
$(x,y,z)$ of the eigenbasis.

Figures \ref{fig:SqCloakrotgrid}b-d show the values of the relative
permeabilities $\mu_x:=\mu_{r}^{11}$ and $\mu_y:=\mu_{r}^{22}$ and
the relative permittivity $\epsilon_z :=\epsilon_{r}^{33}$ in
dependence on the location within the cloak. These three physical
quantities deliver a full description of the propagation behavior of
an electromagnetic wave with a linear polarization vector of the
electric field oriented along the z-direction. The depicted area
corresponds to the green shadowed region in fig.\
\ref{fig:SqCloakTrafo}a. As opposed to $\mu_y$ and $\epsilon_z$, the
value of $\mu_x$ diverges at the boundary of the inner square of the
cloak. However, at a distance of about 1.6~mm from the inner
boundary of the square cloak, $\mu_x$ already approaches finite
values below 35 as determined along a straight intersection line
parallel to the y-axis at $x=3.16$~cm. Please note, that fig.\
\ref{fig:SqCloakrotgrid}b only displays the permeability $\mu_x$ for
$x\geq3.16$~cm and thus does not show the divergence of the relative
permeability towards the inner boundary of the cloak. However,
assuming a typical unit cell size of 3.3~mm for a fabricated
effective medium at a working frequency of 8.5~GHz, the value of
$\mu_x$ at the midpoint of the unit cells at the inner boundary of
the cloak is between 19 and 35, so that an implementation of such a
material is still possible. Although the effective $\mu_x$ is
inaccurate in the vicinity of the boundary of the inner square, it
can be shown, that the performance of the implemented device is not
affected by this fact, which is out of the scope of this paper.
\subsection{Cylindrical Concentrator}
\label{subsec:Trafo_Concentrator}
Due to its cylindrical symmetry, it is convenient to describe the
transformation equations in a cylindrical coordinate system. In this
context, it is necessary to consider the transformation from
cartesian to cylindrical coordinates for an isotropic medium with
permittivity $\epsilon$ and permeability $\mu$. With the
transformation
\begin{eqnarray}
r'(x_1,x_2,x_3) &=& \sqrt{x_{1}^{2}+x_{2}^{2}} \\
\phi'(x_1,x_2,x_3) &=& \arctan{\left( \frac{x_2}{x_1}\right)} \\
x_{3}'(x_1,x_2,x_3) &=& x_3
\end{eqnarray}
and equations (\ref{formula:epsilon}) and (\ref{formula:mu}) one
obtains
\begin{eqnarray}
\eta^{i'j'} = \left( \begin{array}{ccc} r' \eta & 0 & 0
\\ 0 & \frac{\eta}{r'} & 0 \\ 0 & 0 & r' \eta \end{array} \right)
\quad \textrm{with} \quad \eta=\epsilon,\mu \label{formula:epscyl}
\end{eqnarray}
At this point, the reader should be aware, that the metric tensor of
the transformed space is
\begin{eqnarray}
g_{i'j'} = \left( \begin{array}{ccc} 1 & 0 & 0
\\ 0 & r'^{2} & 0 \\ 0 & 0 & 1 \end{array} \right)
\end{eqnarray}
so that (\ref{formula:epscyl}) necessarily still represents the
material properties of an isotropic medium.

The transformation equations for the optical design of the
cylindrical concentrator are denoted as
\begin{eqnarray}
r''(r',\phi',x_{3}') = \nonumber
\end{eqnarray}
\begin{eqnarray}
\begin{cases} \frac{R_1}{R_2}r' &
0\leq r' \leq R_2 \\
\frac{R_3-R_1}{R_3-R_2}r'-\frac{R_2-R_1}{R_3-R_2}R_3 & R_2< r' \leq
R_3 \end{cases} \label{formula:rConc}
\end{eqnarray}
\begin{eqnarray}
\phi''(r',\phi',x_{3}') &=&  \phi' \quad\quad\quad\quad  0\leq\phi'<2\pi  \label{formula:phiConc}\\
x_{3}''(r',\phi',x_{3}') &=& x_{3}' \quad\quad\quad\quad -\infty<
x_{3}'<\infty \label{formula:zConc}
\end{eqnarray}
with the corresponding Jacobi tensors and determinants
\begin{equation}
A^{i'}_{i} =
\begin{cases}
\left( \begin{array}{ccc} \frac{R_1}{R_2} & 0 & 0 \\
0 & 1 & 0 \\
0 & 0 & 1 \end{array} \right) & 0\leq r' \leq R_2 \\
\left( \begin{array}{ccc} \frac{R_3-R_1}{R_3-R_2} & 0 & 0 \\
0 & 1 & 0 \\
0 & 0 & 1 \end{array} \right) & R_2< r' \leq R_3
\label{formula:JacConc}
\end{cases}
\end{equation}
\begin{equation}
det(A^{i'}_{i}) =
\begin{cases}
 \frac{R_1}{R_2} & 0\leq r' \leq R_2 \\
 \frac{R_3-R_1}{R_3-R_2} & R_2< r' \leq R_3
 \label{formula:DetJacConc}
\end{cases}
\end{equation}
\begin{figure}
\centering
\includegraphics[width=3in]{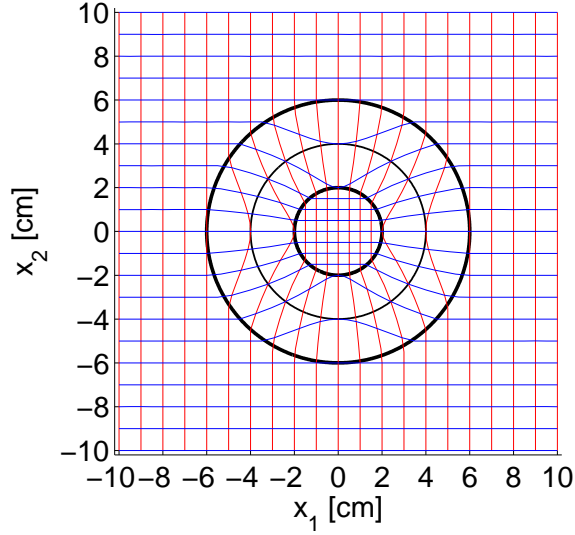}
\caption{\label{fig:ConcGrid} Visualization of the space
transformation expressed in
(\ref{formula:rConc})-(\ref{formula:zConc}). Space is compressed
into a cylindrical region with radius $R_1=2$~cm (black inner
circle) at the expense of an expansion of space between $R_1$ and
$R_3=6$~cm (black outer circle). The displayed intermediate circle
is located at $R_2=4$~cm.}
\end{figure}
The space transformation is visualized in fig.\ \ref{fig:ConcGrid}.
Space is compressed into a cylindrical region with radius R1 at the
expense of an expansion of space between R1 and R3. The
transformation is continuous to free space at R3. Inserting
(\ref{formula:epscyl}), (\ref{formula:JacConc}) and
(\ref{formula:DetJacConc}) into
(\ref{formula:epsilon}-\ref{formula:mu}) and renormalizing
(\ref{formula:epscyl}) by requiring ($\eta^{1'1'}\mapsto
\eta^{1'1'}/(r'\eta), \eta^{2'2'}\mapsto \eta^{2'2'}r'/\eta,
\eta^{3'3'}\mapsto \eta^{3'3'}/(r'\eta)$) to conveniently describe
the relative material properties of free space as
$\eta_r^{i'j'}=\delta^{i'j'}$ in cylindrical coordinates, one
obtains with help of the inverse transformations of
(\ref{formula:rConc}-\ref{formula:zConc}) the relative permittivity
and permeability tensors, expressed in the coordinates
$(r'',\phi'',z'')$ as
\begin{eqnarray}
\epsilon_r^{i''j''} = \mu_r^{i''j''} = \nonumber
\end{eqnarray}
\begin{equation}
\begin{cases}
\left( \begin{array}{ccc} 1 & 0 & 0 \\
0 & 1 & 0 \\
0 & 0 & \left( \frac{R_2}{R_1} \right)^2 \end{array} \right) & 0\leq r'' \leq R_1 \\
\left( \begin{array}{ccc} \eta_r & 0 & 0 \\
0 & \left( \eta_r \right) ^{-1} & 0 \\
0 & 0 & \left( \frac{f}{h}\right) ^2 \eta_r
\end{array} \right) & R_1< r'' \leq R_3 \label{formula:epsmuConc}
\end{cases}
\end{equation}
with
\begin{equation}
\eta_r = \frac{e}{f} \frac{R_3}{r''}+1
\end{equation}
\begin{equation}
 e:=R_2-R_1
\text{, } f:=R_3-R_2 \text{, } h:=R_3-R_1
\end{equation}
Again, in the material interpretation, (\ref{formula:epsmuConc})
represents the material properties of the cylindric concentrator in
the original space $(r',\phi',z')$.
\begin{figure}
\begin{center}
\includegraphics[width = 3.2in]{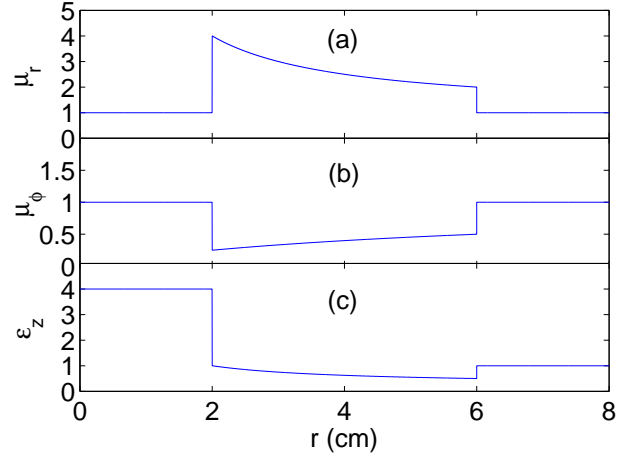}
\caption{\label{fig:Concparam} Radial dependence of the material
properties of a cylindric concentrator with an inner radius
$R_1=2$~cm and an outer radius $R_2=6$~cm. (a) radial permeability
component $\mu_r$, (b) azimuthal permeability component
$\mu_{\phi}$, (c) z-component $\epsilon_z$ of the permittivity }
\end{center}
\end{figure}

Fig.\ \ref{fig:Concparam} shows the radial and azimuthal components
$\mu_r:=\mu_r^{1'1'}$ and $\mu_{\phi}:=\mu_r^{2'2'}$ and the
z-component $\epsilon_z:=\epsilon_r^{3'3'}$ in dependence on the
radial position within the concentrator, assuming an interaction
with an electromagnetic wave with a polarization of the electric
field parallel to the z-direction. The experimental implementation
of such a material requires independent control of the local values
of all three parameters. The metamaterial design and fabrication is
part of current research.
\section{Simulation Results and Discussion}
\label{sec:Simresults}
\begin{figure}
\begin{center}
\includegraphics[width = 3in]{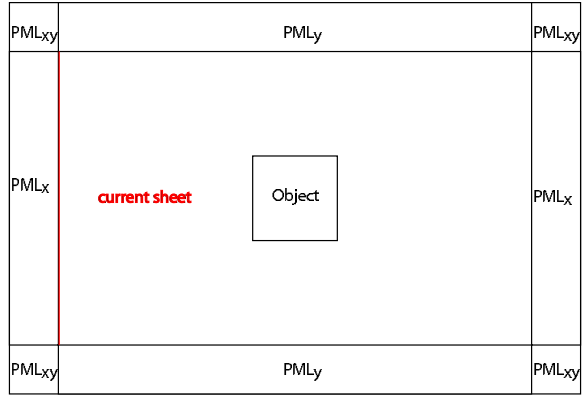}
\caption{\label{fig:CompDomain} Computational domain terminated by
perfectly matched layers with functionality in the x-, y- and
x-y-direction. The TE plane wave is excited by a current sheet.}
\end{center}
\end{figure}

For the full wave electromagnetic simulations a two-dimensional
finite-element solver of the Comsol Multiphysics software package
was used. The computational domain and its boundaries are shown in
fig.\ \ref{fig:CompDomain}. A transverse-electric (TE) plane-wave
was excited by a current sheet. The computational domain was
terminated by perfectly matched layers (PMLs). The current density
distribution was chosen to exponentially decrease towards the
borders of the sheet in the y-direction to prevent interactions with
the PMLs. The calculations did not include absorption in the object.
The finite-element solver required all physical quantities to be
described in cartesian coordinates. For all simulations, a
stationary solver was used. The solver allowed to conveniently
implement the functional dependence of the permittivity and
permeability tensors of the simulated material into the model and
thus to accurately describe and predict the electromagnetic behavior
of the designed medium.
\subsection{Square Cloak}
\label{subsec:Simresults_Sqcloak}
\begin{figure*}
\centering
\includegraphics[width = 6in]{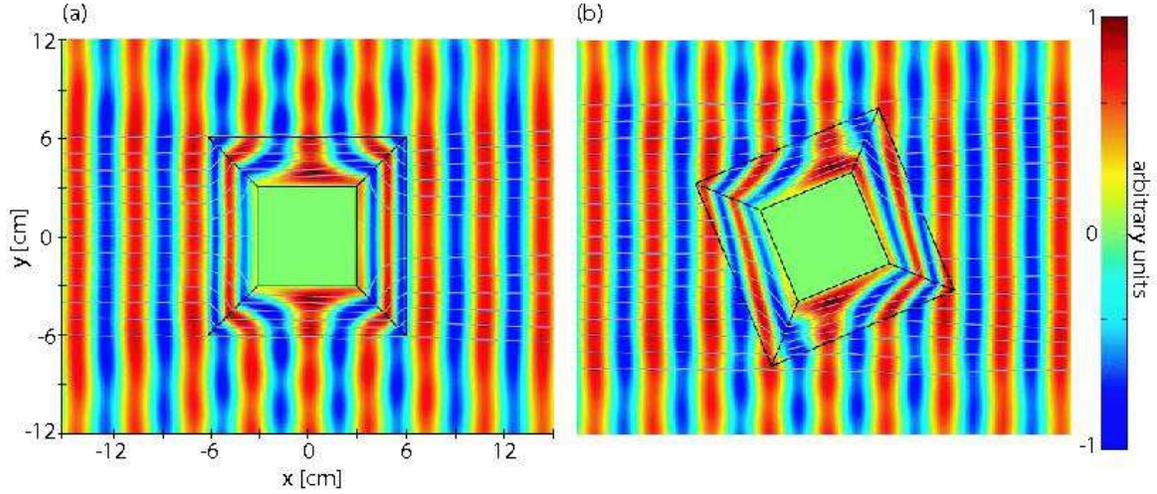}
\caption{\label{fig:Cloak1} Electric field distribution in the
interior and exterior region of the square cloak. The direction of
the power flow is in the positive x-direction. The wave is smoothly
bent around the corners of the square cloak. (a) Phase fronts
parallel to one side of the cloak, (b) Cloak rotated by $\pi/8$ with
respect to the phase fronts of the incoming wave}
\end{figure*}

Fig.\ \ref{fig:Cloak1} shows the results of the two-dimensional
full-wave simulations of a square-shaped cloak. The color map
depicts the spatial distribution of the real part of the
transverse-electric phasor oriented along the z-direction. In
addition, the direction of the power flow is indicated by the grey
lines. The frequency of the TE wave is 8.5~GHz. The sidelengths of
the inner and outer square of the cloak are 6~cm and 12~cm,
respectively.

In fig.\ \ref{fig:Cloak1}a, the phase fronts of the impinging wave
are parallely aligned to one of the sides of the cloak. As can be
seen, the wave is smoothly bent around the cloaked area and the
phase fronts are completely restored when the wave exits the cloak
material. The inhomogeneity and the anisotropy of the cloak medium
are evident as the direction of the power flow and the phase front
normal are not parallel and the angle between the directions changes
locally. In fig.\ \ref{fig:Cloak1}b, the cloak is rotated by an
angle of $\pi/8$ with respect to the phase fronts of the incoming
wave. In this configuration, the phase fronts are no longer parallel
to any side of the square cloak. As before, the phase fronts are
completely restored after propagation through the cloak material and
the inner square is not sensed by the wave. In both cases, the wave
impedances of the cloak medium and free space are exactly matched
and the device is therefore reflectionless.

The square-shaped cloak is an example of a cloak with reduced
symmetry in comparison to a cylindric cloak. In addition, the square
cloak possesses sharp corners. The simulations clearly show, that
the transformation-optical cloak design is not restricted by
cylindric symmetry requirements. In principle, cloaks of arbitrary
shape can be designed by use of the transformation-optical approach.
\subsection{Concentrator}
\label{subsec:Simresults_Concentrator}
\begin{figure*}
\centering
\includegraphics[width = 6in]{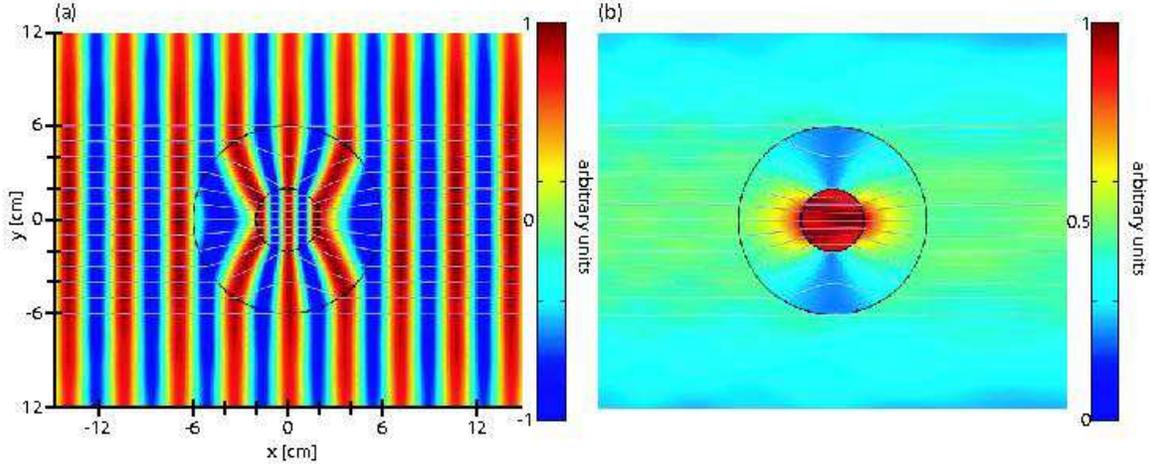}
\caption{\label{fig:Conc} (a) Electric field distribution in the
interior and exterior region of the cylindric concentrator. The
direction of the power flow is in the positive x-direction,
indicated by the grey lines. (b) Normalized power flow distribution.
The power flow is enhanced within the region with radius R1=2~cm by
a factor of 2. Much stronger power flow enhancements can be achieved
by increasing the ratio $R_2/R_1$.}
\end{figure*}
In section \ref{subsec:Trafo_Concentrator}, the material properties
of a concentrator were described in a cylindrical coordinate system.
In cartesian coordinates, the relative permittivity and permeability
tensors can be obtained from (\ref{formula:epsmuConc}) by use of the
general transformation
\begin{eqnarray}
\xi^{i'j'} = \nonumber
\end{eqnarray}
\begin{equation}
\left( \begin{array}{ccc} \frac{1}{r^2}\left( \xi^{11} x_{1}^2+
\xi^{22}x_2^{2} \right) &
(\xi^{11}-\xi^{22})\frac{x_1 x_2}{r^2} & 0 \\
(\xi^{11}-\xi^{22})\frac{x_1 x_2}{r^2} & \frac{1}{r^2}\left(\xi^{22}
x_{1}^2+ \xi^{11}x_2^{2} \right) & 0 \\
0 & 0 & \xi^{33} \end{array}\right) \label{formula:epsmuConccart}
\end{equation}
with $\xi=\epsilon_r\text{, }\mu_r$. The variables $\epsilon_r$ and
$\mu_r$ as functions of the space variables $(x:=x_1, y:=x_2,
z:=x_3)$ are then directly assigned to the concentrator domains.

Fig.\ \ref{fig:Conc}a displays the real part of the phasor of the
electric field for a z-polarized TE wave. The grey lines represent
the direction of the power flow. The free-space frequency of the TE
wave is 8.5~GHz. The outer radius of the concentrator is $R_3 =
6$~cm. As can be seen, the fraction of the plane-wave extending in
the y-direction from $-R_2=-4$~cm to $R_2=4$~cm is completely
focussed by the concentrator into the region with radius $R_1=2$~cm.
Additionally, the fields within the intervals $[-R_3,-R_2)$ and
$(R_2,R_3]$ in the y-direction are focussed to an area with a radius
lying in the interval $(R_1,R_3]$.

Fig.\ \ref{fig:Conc}b illustrates the normalized intensity
distribution of the TE wave. It is obvious, that the field
intensities are strongly enhanced in the inner region of radius
$R_1$ within the concentrator material. The intensity enhancement
factor for the chosen structure, computed as the ratio between the
maximal values of the field intensities outside the circular region
with radius $R_3$ and inside the concentrator region with radius
$R_1$, is 2. Significantly stronger enhancements can be achieved by
increasing the ratio $R_2/R_1$. As can be seen, the enhancement
theoretically diverges to infinity as $R_1$ goes to zero.

Due to the rotational symmetry around the axis perpendicular to the
$x$-$y$-plane, the concentrator focusses waves impinging from
arbitrary directions. The concentrator is reflectionless due to
inherent impedance matching in the transformation-optical design
method. Although metamaterials, which are necessary to implement the
material properties of a concentrator, inherently suffer from
losses, we think, that the concentrator can play an important role
in the harnessing of light in solar cells or similar devices, where
high field intensities are required.
\section{Conclusion}
In conclusion, we have presented the material design of a
square-shaped cloak and an electromagnetic field concentrator based
on form-invariant transformations of Maxwell's equations. The
electromagnetic behavior of the devices was simulated by use of a
two-dimensional finite element solver. In contrary to previous
publications, the simulated cloaking device did not possess a
cylinder-symmetry. The proposed electromagnetic field concentrator
proved to be well suited for the confinement of electromagnetic
energy of waves impinging from arbitrary directions. The two
demonstrated optical devices exemplify the strength of the general
methodology of form-invariant coordinate transformations of
Maxwell's equations for the design of electromagnetic materials with
a well-defined functionality. The technique allows to chose from an
infinite set of allowed transformations and thus provides a powerful
tool for the conception of optical elements with previously
unachievable electromagnetic behavior.

D. Schurig wishes to acknowledge support from the IC Postdoctoral
Research Fellowship program.




\end{document}